\documentclass[10pt]{blois07}
\usepackage{graphicx}
\usepackage{cite,./mcite}
\usepackage{slashed}
\usepackage{amsmath,amssymb,latexsym}

\def\o{\omega}
\newcommand{\I}{\textrm{Im}}
\newcommand{\cA}{{\cal A}}
\newcommand{\cF}{{\cal F}}
\newcommand{\sA}{\slashed{A}}

\newcommand{\mbf}[1]{\mbox{\boldmath $#1$}}
\newcommand{\bk}{\mbf{k}}
\newcommand{\bq}{\mbf{q}}

\newcommand{\bb}{\mbf{b}}

\begin{document}
\title{Multiple interactions and AGK rules in pQCD}
\author{M. Salvadore$^1$\protect\footnote{~~Talk presented at EDS07}~~,
J. Bartels$^1$, G. P. Vacca$^2$}
\institute{$^1$II. Institut f\"ur Theoretische Physik, Universit\"at Hamburg,
Luruper Chaussee 149, 22761 Hamburg, Germany,\\
$^2$Dipartimento di Fisica - Universit\`a di Bologna and INFN -
Sezione di Bologna, via Irnerio 46, 40126 Bologna, Italy}
\maketitle
\begin{abstract}
We review some aspects of multiple interactions in High Energy QCD;
we discuss in particular AGK rules and present some results concerning
multiple interactions in the context of jet production.
\end{abstract}

\section{Introduction}
\label{sec:intro}
Many years ago Abramovsky, Gribov and Kancheli in their pioneering
paper \cite{Abramovsky:1973fm} have pointed out that, for high energy
hadron-hadron scattering, multiple exchange of pomerons leads to observable
effects in multiparticle final states.
Multi pomeron exchange induces indeed fluctuations in the rapidity
densities of the produced particles; concerning the multiple inclusive
production of jets, they cause long range rapidity correlations.
Nowdays it has become evident that multiple interactions play a substantial
role in determining the behaviour of high energy scattering.
As in-depth studies of DIS at HERA in the small x region
and jet physics at Tevatron have shown, diffractive events represent
a substantial fraction of the total cross section.

The advent of LHC will open up a much wider kinematical window with respect
to any other hadron collider which as been available so far.
Needless to say, the challenging measurements aimed at the discovery of
physics beyond the Standard Model require an extremely precise understanding
of the background physics. In particular, it is needed to assess the
effects introduced by multiple interactions. There are kinematical regions
where the power suppression due to the ``higher twist'' nature of these
effect is expected to be compensated.
If a jet is produced close to the forward direction for example, one of
the colliding hadrons PDF is probed in the region of very small
longitudinal fraction, where the dominant gluon density undergo a step
rise of the type $\propto (1/x)^\lambda$, $x \to 0$, $\lambda > 0$.
Here is where the mellow concepts developed in the pre QCD era becomes
topical again.

As far as the theoretical motivations for considering multiple interactions
are concerned, it is well known that they are expected to unitarize
cross sections. The resummation of Leading Logarithms (LL) $\log (1/x)$
in pQCD results in the perturbative BFKL pomeron \cite{Kuraev:1976ge,
Kuraev:1977fs,Balitsky:1978ic}, which violated the unitarity constraint
expressed by the Froissart bound, $\sigma_\textrm{tot} \le \log^2(s)$,
$s \to \infty$. Finding a systematic way of including a minimal subset
of subleading corrections in order to restore unitarity has been subject
to intensive research in the past decade.

In this talk we discuss AGK rules in the context of
pQCD and some recent results regarding multiple interaction effects in
inclusive jet production.
The relevant papers are \cite{Bartels:2005wa,inprep}.

\section{Review of AGK rules in Regge theory}
\label{sec:revAGKRegge}
In their simplest version, AGK cutting rules are nothing but
a statement about how the $s$-channel unitarity is encoded
in reggeon diagrams. In standard perturbation theory
the corresponding tool is the set of Cutkosky rules \cite{Cutkosky:1960sp},
which tells us how to build up the discontinuity of an amplitude
at a particular order in the coupling, summing up all the possible
cut diagrams contributing to the amplitude. Symbolically, we
could write
\begin{equation}
  \label{eq:cutkosky}
  2\I \cA=\sum_\textrm{cuts}\sA,
\end{equation}
where $\sA$ indicates generically any cut version of the amplitude $\cA$.
Such an approach is clearly unfeasible when considering reggeon diagrams,
since their being already the sum of a full series in the coupling
implies that we should consider an infinite set of cut amplitudes
already at the simplest level beyond a single ladder.
Each term in the sum of multi-ladder diagrams contains a number of
possible cuts quickly growing with the number of the rungs in each ladder.

The fundamental observation of AGK was that cut reggeon diagrams where
the cut cross a side line of the ladder are strongly suppressed compared to
diagrams where the reggeons are cut or uncut completely.
The natural question arising from this observation is whether
the latter set of diagrams is alone sufficient to reconstruct
the full discontinuity; the striking result pointed out in the
AGK papers is the affirmative answer to this question.
Let us then review their arguments.

The starting point of the AGK analysis is the Sommerfeld-Watson
representation of the elastic scattering amplitude $\cA$:
\begin{equation}
  \label{eq:SWrep}
  \cA(s,t) = 
  \int \!\! \frac{d\o}{2i} \xi(\o) s^{1+\o} \cF(\o,t),
  \qquad\qquad
  \xi(\o)= \frac{\tau-e^{-i \pi \o}}{\sin \pi \o} \, .
\end{equation}
The signature function can be found in \cite{Bartels:2005wa}.
The (real-valued) partial wave $\cF(\o,t)$ has singularities in the 
complex $\o$-plane, and the multi-Regge exchange corresponds to a 
particular branch cut.
As we have already pointed out,
the central goal of the AGK analysis is the decomposition of the contribution
of the $n$-reggeon cut in terms of $s$-channel intermediate states.
The absorptive part of the amplitude will consist of several different 
contributions: each piece belongs to a particular energy cut line,
and there are several different ways of drawing 
such energy-cutting lines. Each of them
belongs to a particular set of $s$-channel intermediate states. 
For example, a cutting line between reggeons,
\begin{center}
\includegraphics[width=2.5cm]{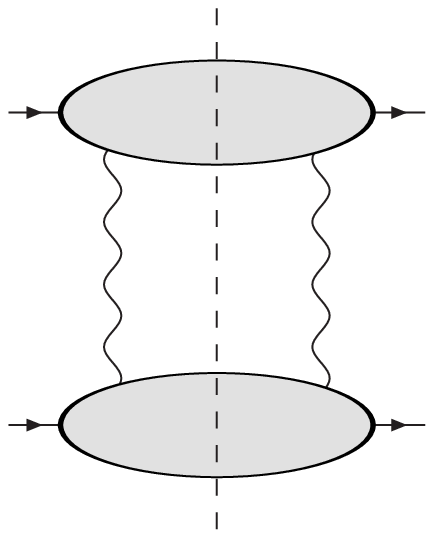}
\end{center}
belongs to double diffractive production on both sides of the cut:
there is a rapidity gap between what is inside the upper blob and the 
lower blob. When relating this contribution with the full diagram,
one requires a `cut version' of the reggeon
particle couplings $N_n$ (represented as grey blobs in the figures).
Similarly, the cut through a reggeon,
\begin{center}
\includegraphics[width=2.5cm]{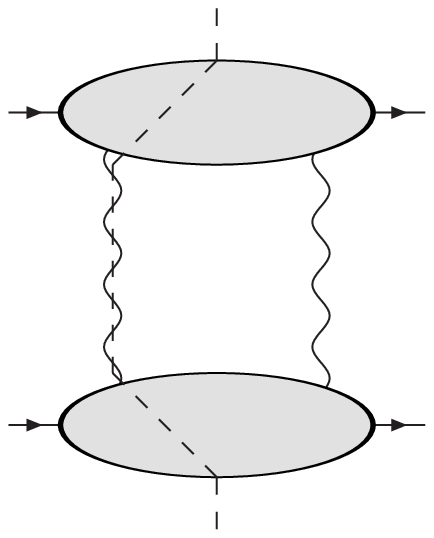}
\end{center}
corresponds to a so-called
multiperipheral intermediate state, and another cut version of the 
particle-reggeon coupling appears.
The basis of the AGK analysis is the observation that, under very
general assumptions for the underlying dynamical theory,
these couplings are fully symmetric under the exchange
of reggeons, and all their cut versions are identical.
This property then allows to find simple relations between
the different cut contributions, and to derive a set of counting rules.

Due to the analytic structure of the partial wave $\cF(\o,t)$ in the
$s$-channel physical region, which is given by a set of cuts and poles
on the real axis on the right side of a fixed point $\o=\o_0$, the amplitude
$\cA$ can be decomposed into the sum of the different contributions
due to the various singularities,
\begin{equation}
  \label{eq:AdecompinAn}
  \cA(s,t) = \sum_{n=1}^{\infty} \cA_n(s,t),
\end{equation}
where $\cA_n(s,t)$ is the contribution due to the $n$-reggeon branch point.
A further decomposition of $\I \cA_n$ was found by AGK
by observing that \emph{the complete result is
obtained by summing up just the diagrams where the reggeons
are cut or uncut completely}, therefore neglecting all the
(multitude of) diagrams where the cutting line breaks up at least
one of the reggeons. For the $n$-reggeon cut there are $n+1$ of
such contributions (any number of cut reggeon from 0 to n),
and one ends up with
\begin{equation}
  \label{eq:ImAsum}
  \boxed{2 \I \cA(s,t) = \sum_{n=1}^{\infty} \sum_{k=0}^{n} \sA_{nk}(s,t).}
\end{equation}
Comparing this last equation with \eqref{eq:cutkosky},
it is immediately clear that we have struck a big deal:
in building up the imaginary part of the amplitude, we have got
rid of most of the cuts on the r.h.s. of \eqref{eq:cutkosky},
only those in \eqref{eq:ImAsum} being left.

The simplest case, the two-Pomeron
exchange, has the three contributions:
\begin{center}
\includegraphics[width=8cm]{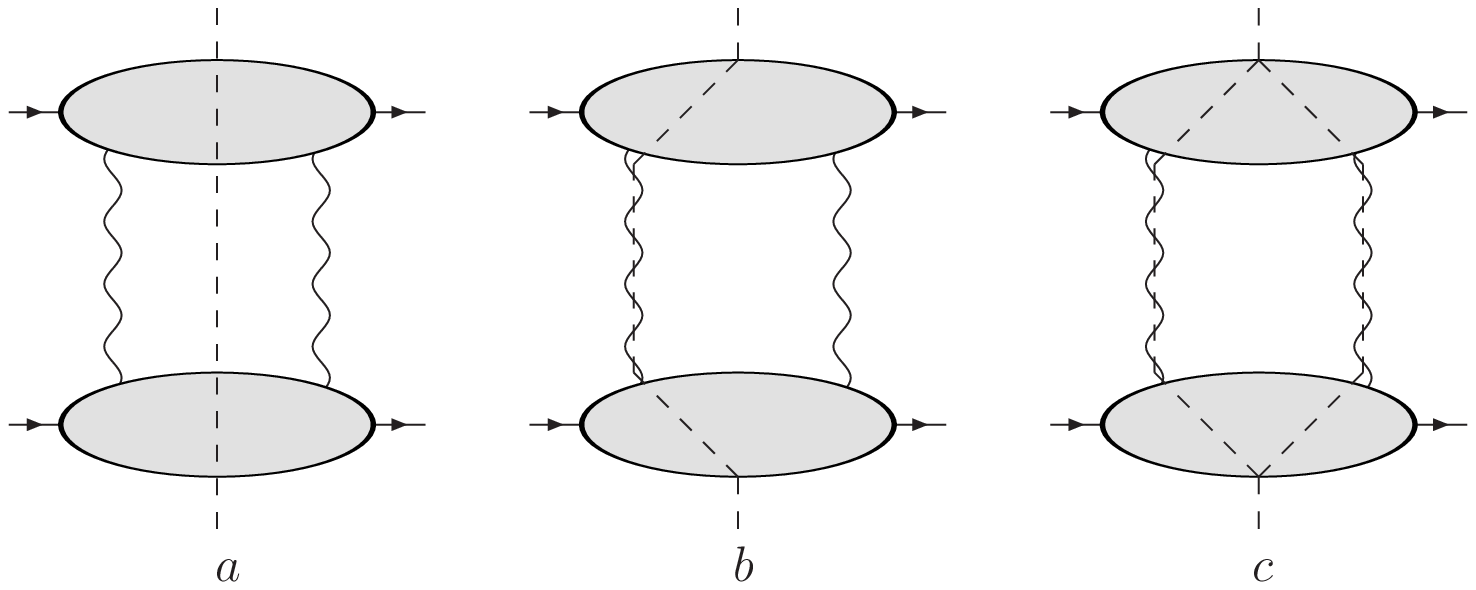}
\end{center}
(a) in the diffractive cut all the pomerons are left uncut,
and there is a rapidity gap between the fragmentation regions of the
two particles;
(b) in the single multiplicity cut only one pomeron has been cut;
(c) when both pomerons are cut the multiplicity of particles is
doubled with respect to the previous case.
In this case one obtains the well known result that the different contributions
are in the proportion
\begin{equation}
  \label{eq:prop}
  \sA_{20}:\sA_{21}:\sA_{22} = 1:-4:2.
\end{equation}
Unfortunately AGK constraints can be formulated only for a very restricted
class of interaction vertices (in particular,
for the $1 \to n$ Pomeron vertex).
For the general case (for example, for the $2 \to 2$ vertex)
this is not the case; only explicit models,
e.g. calculations in pQCD as the one discussed in section \ref{sec:jet},
can provide further information.

Another remarkable result stems from the AGK analysis:
for the $n$-particle inclusive cross section,
large classes of multi-pomeron corrections cancel.
For the single inclusive case
all multi-Pomeron exchanges across the produced particle cancel,
\begin{equation}
  \label{eq:AGKcancellation1jetPic}
  \parbox{60pt}{\includegraphics[width=60pt]{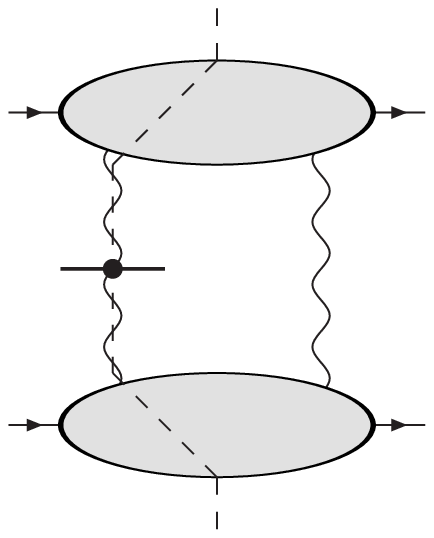}}
  +
  \parbox{60pt}{\includegraphics[width=60pt]{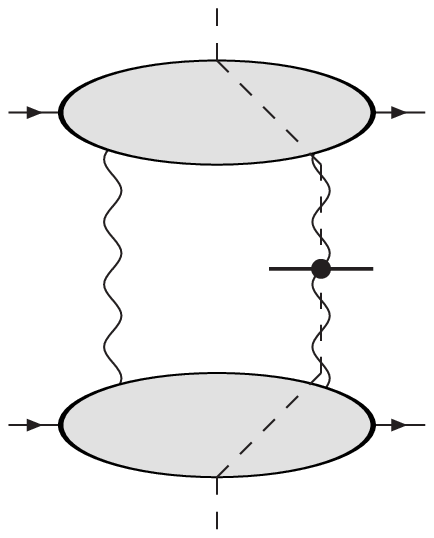}}
  +
  \parbox{60pt}{\includegraphics[width=60pt]{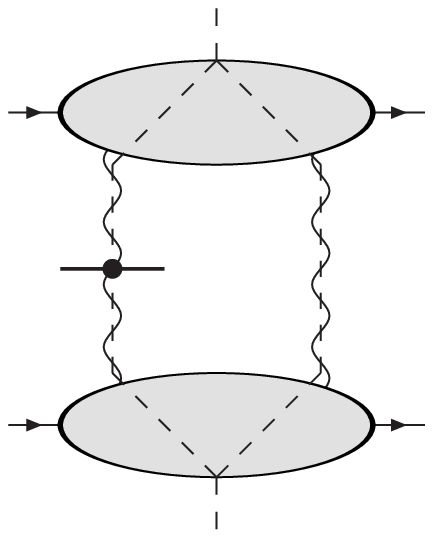}}
  = 0
\end{equation}
We are now ready to discuss the novel features arising in QCD.

\section{AGK rules in pQCD}
\label{sec:AGKpQCD}

From this brief review it follows that the central task of 
performing the AGK analysis in pQCD requires the computation and study of 
the coupling functions $N_n$. The simplest task is the study of the 
two-Pomeron exchange. Since the BFKL Pomeron 
is a composite state of two reggeized gluons, we have to start from the 
exchange of four reggeized gluons.

In the simple perturbative situation where the external particles
are photons, the computation of these couplings (which are denote $D_n$)
was performed in \cite{Bartels:1994jj}. The two pomerons exchange
contribution is  encoded in the amplitude $D_4$.
The most peculiar fact about $D_4$ is its decomposition into two pieces:
\begin{equation}
  \label{eq:Z4decompAGK}
  D_4 = D_4^{~I} + D_4^{~R}.
\end{equation}
The first term, $D_4^{~I}$, is completely symmetric under the exchange of any
two gluons, whereas the second one, $D_4^{~R}$, is a sum of antisymmetric
terms which, as a result of bootstrap properties, can be expressed
in terms of two-gluon amplitudes, $D_2$.
Under the exchange of the two reggeized gluons,
$D_2$ is symmetric. It is only after this decomposition has been
performed, and we have arrived at reggeon particle couplings with `good'
properties, that we can start with the AGK analysis.

Starting from these functions $D_4^{~I}$ and $D_4^{~R}$, the investigation in
\cite{Bartels:1996hw} has shown in some detail how the AGK counting
rules work in pQCD: the analysis has to be done
seperately for $D_4^{~I}$ and $D_4^{~R}$. For the former piece, we obtain
the counting arguments for the Pomerons (which is even-signatured)
given by AGK;
here the essential ingredient is the complete symmetry of $D_4^{~I}$ under the
permutation of reggeized gluons.
In the latter piece, the odd-signature reggeizing gluons lead to
counting rules which are slightly different from those of the even signature
Pomeron: once the bootstrap properties have been invoked and $D_4^{~R}$ is
expressed in terms of $D_2$ functions, cutting lines through the reggeized
gluon appear. Since it carries negative signature, the relative weight
between cut and uncut reggeon is different from the Pomeron.

In the light of these facts, we will now attempt to use of the pQCD
cutting rules in a nonperturbative environment (e.g.multi-ladder exchanges
in $pp$ scattering). Basic ingredients are the
nonperturbative couplings of $n$ reggeized gluons to the proton.
In order to justify the use of pQCD we need a hard scattering subprocess:
we will assume that all reggeized gluons are connected to some hard
scattering subprocess; consequently, each gluon line will have its
transverse momentum in the kinematic region where the use of pQCD can be
justified. Since AGK applies to the high energy limit
(i.e. the small-$x$ region), all $t$-channel gluons are reggeized.
Based upon the analysis in pQCD,
we now formulate a few general conditions which the nonperturbative
couplings $N_n$ have to satisfy in order to get the usual AGK counting rules:
\begin{quote} \em
(i) they are symmetric under the simultaneous exchange of momenta
and color;\\
(ii) cut and uncut vertices are identical, independently where the
cut line enters.
\end{quote}
Whenever these two properties are satisfied, it can be proved that the
$n$-reggeized gluon cut satisfies a similar set of counting rules as the
original ones found by AGK, but note that, in contrast to the 
discussion above, in the case of reggeized gluons we do not need to consider
cutting lines inside the reggeized gluons: compared to an uncut gluon, 
a cut gluon line is suppressed in order $\alpha_s$. The multiplicity
of the final state arises due to the $s$-channel gluons mediating the
interactions between the reggeized ones.

A simple (oversimplified) model for the coupling $N_n$ correspond
to eikonal couplings:
\begin{equation}
  \label{eq:eikonalNn}
  N_{2n}= \Phi(1,2)\Phi(3,4)\ldots\Phi(2n-1,2n)+\textrm{permutations}
\end{equation}
Squaring two of these couplings and taking the large $N_c$ limit one obtains
that the multiplicity-$k$ contribution to the total cross section is
\begin{equation}
  \label{eq:multKsigma}
  \sigma_k = \int d^2\bk~ e^{i \bb \cdot \bq} P_k(s,\bb) \, ,
\end{equation}
where $P_k$ is a Poissonian distribution,
\begin{equation}
  \label{eq:Pk}
  P_k(s,\bb) = \frac{\Omega(s,\bb)^k}{k!} e^{-\Omega(s,\bb)} \, ,
\end{equation}
which is interpreted as the probability to have $k$ cut pomerons at fixed
impact parameter $\bb$ and total energy $\sqrt{s}$.

\section{Multiple interactions in jet production}
\label{sec:jet}
The other remarkable result of AGK, the destructive interference leading
to the cancellations of diagrams for jet production as the one represented 
in eq. \eqref{eq:AGKcancellation1jetPic}, has also a counterpart in QCD.
The analogous result in QCD reads
\begin{equation}
  \label{eq:AGKcancellation1jetQCD}
  \sum
  \parbox{60pt}{\includegraphics[width=60pt]{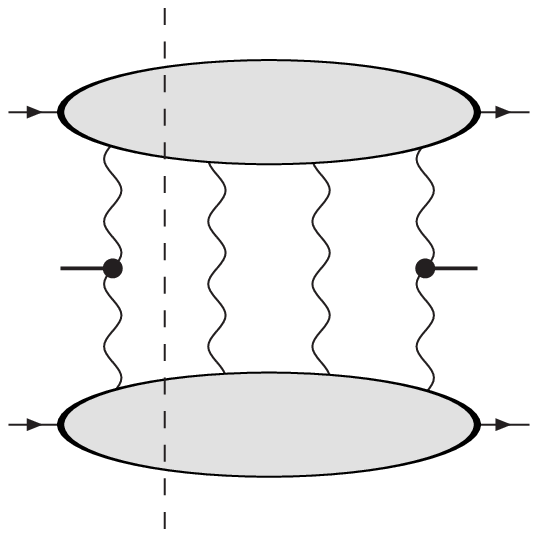}}
  +
  \parbox{60pt}{\includegraphics[width=60pt]{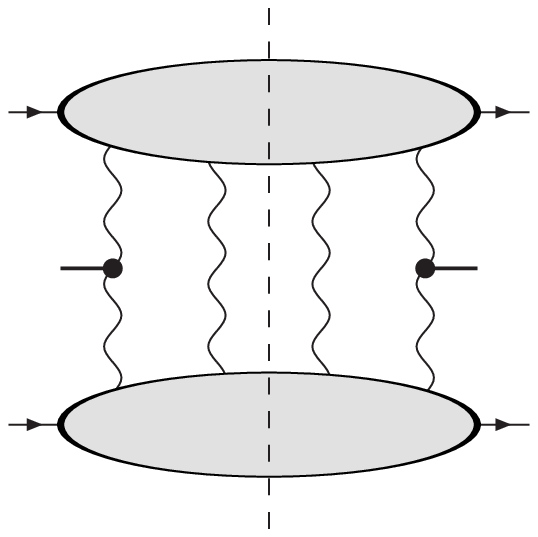}}
  +
  \parbox{60pt}{\includegraphics[width=60pt]{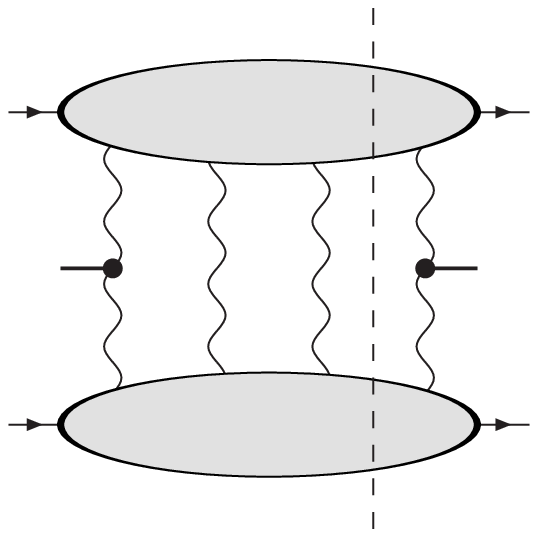}}
  = 0 \, ,
\end{equation}
where the sum is over all the possible ways to attach the jet (gluon)
on each side of the cut.
Such interference take place only due to the summation over all possible
final states and integration over the phase space. Any given final state
(underlying event), would provide by itself a non-vanishing contribution.
The result expressed by eq. \eqref{eq:AGKcancellation1jetQCD} is quite general;
it holds indeed for an arbitrary number of reggeized gluons and jets
(see \cite{Bartels:2005wa}).

What is left after these cancellation have been exploited are production
vertices for the reggeized gluon interactions. such vertices are model
dependent, and must be computed from the underlying theory.
The first step in this direction has been taken in \cite{inprep}, where
the three cuts of the two-to-four reggeized gluon vertex have been computed.
The techniques used are similar to those exploited in \cite{Bartels:1994jj}
for the computation of the two-to-four inclusive vertex (triple pomeron
vertex). One writes down a set of coupled evolution equations for
the particle-reggeized gluon couplings, where the evolution is given
by the exchange of $s$-channel gluons, and the virtual corrections are taken
into account by properly including the gluon Regge trajectories.
In the case of single jet production, the kinematics of a gluon is kept
fixed. Reshuffling such equation and using the bootstrap property, one
obtains the factorization expressed by eq. \eqref{eq:Z4decompAGK} for
the inclusive case.

In the single jet production, the coherence leading to the simple
factorization \eqref{eq:Z4decompAGK} is partially broken by the
missing integration over the phase space of the produced jet.
In \cite{inprep} was observed that is possible to obtain gauge invariant
objects by identifying antipodal jets: one gives up the distinction between
jets emitted in opposite directions in the transverse plane.
If doing so, it is possible to factorize the amplitude as a sum of gauge
invariant pieces. Explicitely one gets in a pictorial form
\begin{eqnarray}
  \label{eq:iZ4-total-pic}
    \begin{minipage}{1.5cm}
      \includegraphics[width=2cm]{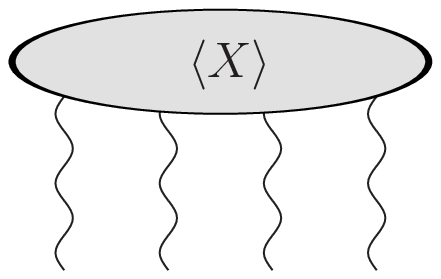}
    \end{minipage} &=&
    \sum \bigg(
    \begin{minipage}{1.5cm}
      \includegraphics[width=1.5cm]{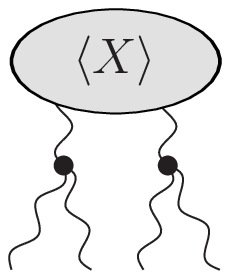}
    \end{minipage} +
    \begin{minipage}{1.5cm}
      \includegraphics[width=1.5cm]{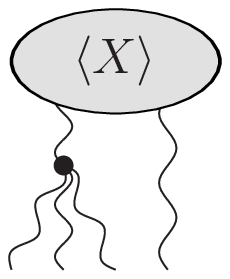}
    \end{minipage} +
    \begin{minipage}{1.5cm}
      \includegraphics[width=1.5cm]{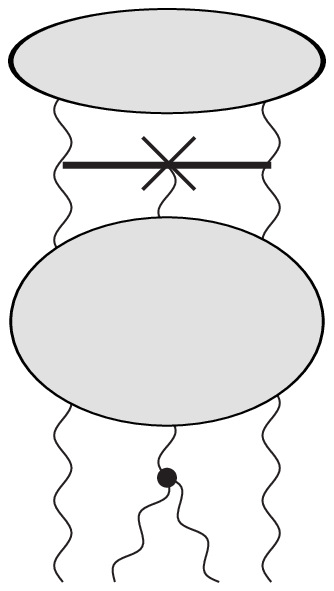}
    \end{minipage} +
    \begin{minipage}{2cm}
      \includegraphics[width=2cm]{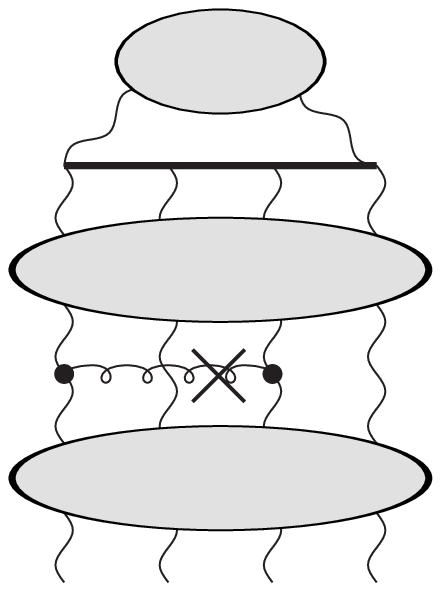}
    \end{minipage}
    \bigg) + \nonumber\\&&+
    \begin{minipage}{2cm}
      \includegraphics[width=2cm]{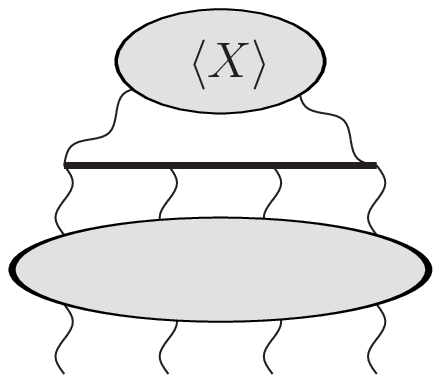}
    \end{minipage} +
    \begin{minipage}{2cm}
      \includegraphics[width=2cm]{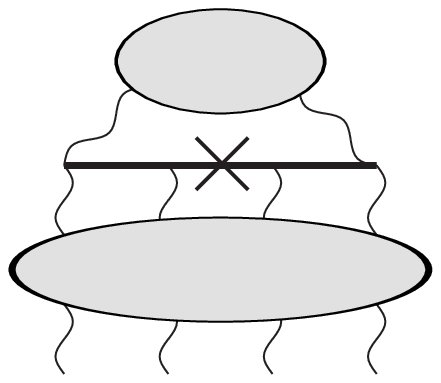}
    \end{minipage} +
    \begin{minipage}{2cm}
      \includegraphics[width=2cm]{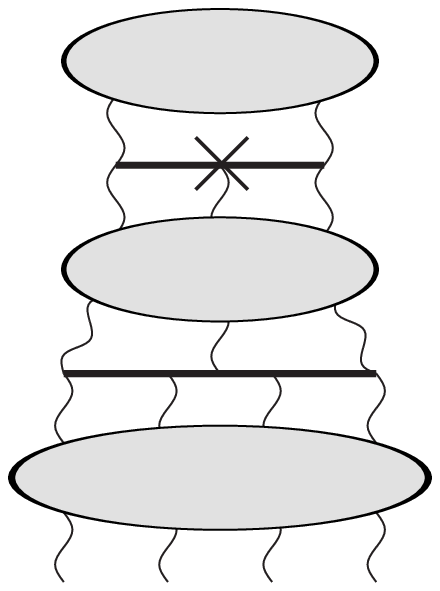}
    \end{minipage} \, .
\end{eqnarray}
The grey blob on the l.h.s. represents the full particle-reggeized gluons
coupling with a gluon fixed. On the r.h.s. one observes the various gauge
invariant terms contributing to such coupling. The first three
represent ``reggeized'' terms: they describe the exchange of less then four
gluons, some of which are ``composite'' (they contain corrections beyond
to the LL reggeized gluon). The fourth term does not contribute to the
cross section thanks to the AGK argument expressed by equation
\eqref{eq:AGKcancellation1jetQCD}.
The first term in the last line is very simple: the jet is emitted inside
the pomeron attached to the external particles, and subsequently the
pomeron decays into a four reggeized gluon state via the standard
two-to-four vertex. The last two pieces contain new ingredients.
In the first appear for the first time the production vertices for
the two-to-four transition with the emission of one jet. There are
three variants of such vertex, depending where the $s$-channel cut passes.
In the last piece the new objects are a universal two-to-three production
vertex, and three versions of a three-to-four inclusive vertex.
More details and explicit expression can be found in \cite{inprep}.

\section{Concluding remarks}
\label{sec:conclusion}
We have reviewed some aspects of multiple interactions in pQCD, in
the context of the total cross section and associated multiplicity
distributions (AGK), and of inclusive jet production. We have shown
that similar cutting rules as those first obtained in the framework
of soft pomeron Regge theory emerge in pQCD as well.
We have also presented new vertices for the inclusive production of a jet
across the transition between two and three or four reggeized gluons.
The full particle-four-gluons coupling has been decomposed in a sum
of gauge invariant pieces, and each of them has been computed explicitely.
Explicit expressions for all the new vertices are computed and presented
in \cite{inprep}.

\begin{footnotesize}
\bibliographystyle{blois07} 
{\raggedright
\bibliography{blois07}
}
\end{footnotesize}
\end{document}